\documentstyle[preprint,tighten,aps]{revtex}
\begin{document}
	\makeatletter
        \@addtoreset{equation}{section}
        \makeatother
        
\renewcommand{\theequation}{\thesection.\arabic{equation}}

        \title{Hausdorff dimension of a quantum string}
	
	\author{ S.Ansoldi\footnote{E-mail address:
         ansoldi@trieste.infn.it}}
        \address
        { Dipartimento di Fisica Teorica dell'Universit\`a,\\
         Strada Costiera 11, 34014-Trieste}                                
	
	\author{ A.Aurilia\footnote{E-mail address:
         aaurilia@csupomona.edu}}
        \address{ Department of Physics, California State 
Polytechnic
	University\\ Pomona, CA 91768, USA,}
	
        \author{ E.Spallucci\footnote{E-mail address:
         spallucci@vstst0.ts.infn.it}}
       \address {\it Dipartimento di Fisica Teorica 
dell'Universit\`a,\\
         Strada Costiera 11, 34014-Trieste, Italy,\\
         Istituto Nazionale di Fisica Nucleare, Sezione di 
Trieste,\\
         Strada Costiera 11, 34014-Trieste, Italy,}
	\maketitle

\begin{abstract}
In the path integral formulation of quantum mechanics, Feynman 
and Hibbs noted that the trajectory of a particle is continuous but
nowhere differentiable. We extend this result to the quantum 
mechanical path of a relativistic string and find that the 
~``trajectory"~, in this case, is a fractal surface with Hausdorff 
dimension three.  Depending on the resolution of the detecting 
apparatus, the extra dimension is perceived as ~``fuzziness"~ of the 
string world-surface. We give an interpretation of this phenomenon 
in terms of a new form of the uncertainty principle for strings, and 
study the transition from the smooth to the fractal phase.
	\end{abstract}
	\newpage

	\section{Introduction and Synopsis}

	A classical string is a one--dimensional, spatially 
extended 
object, so that its timelike orbit in spacetime is described 
by a 
smooth, two--dimensional manifold. However, since the advent 
of 
quantum theory and general relativity, the notion of 
spacetime as a 
preexisting manifold in which physical events take place, is 
undergoing a process of radical revision. Thus, reflecting on 
those 
two major revolutions in physics of this century, Edward 
Witten 
writes\cite{witten}, ~``{\it Contemporary developments in 
theoretical 
physics suggest that another revolution may be in progress,  
through 
which a new source of ~``fuzziness"~ may enter physics, and 
spacetime itself may be reinterpreted as an approximate, 
derived 
concept.}"~The new source of fuzziness comes from string 
theory, 
specifically from the introduction of the new fundamental 
constant, 
($\alpha'$), which determines the tension of the string. 
Thus, at 
scales comparable to $(\alpha')^{1/2}$, spacetime becomes 
fuzzy, even 
in the absence of conventional quantum effects ($h=0$). While 
the 
exact nature of this  fuzziness is unclear, it manifests 
itself in a new 
form of Heisenberg's principle, which now depends on both 
$\alpha'$ 
and $h$. Thus, in Witten's words, while~``{\it a proper 
theoretical 
framework for the [new] uncertainty principle has not yet 
emerged,........the natural framework of the [string] theory 
may 
eventually prove to be inherently quantum mechanical.}"~\\ 
The essence of the above remarks, at least in our 
interpretation, is 
that there may exist different degrees of fuzziness in the 
{\it 
making} of spacetime, which set in at various scales of 
length, or 
energy, depending on the nature and resolution of the 
Heisenberg 
microscope used to probe its structure. In other words, {\it 
spacetime 
becomes a sort of dynamical variable, responding to quantum 
mechanical resolution just as, in general relativity, it 
responds to 
mass--energy}. The response of spacetime to mass--energy is 
curvature. Its response to resolution seems to be 
~``fractalization"~. This, in a nutshell, is the central 
thesis of this 
paper.\\
Admittedly, in the above discussion, the term ~``fuzziness"~ 
is loosely 
defined, and the primary aim of this paper is to suggest a 
precise 
measure of the degree of fuzziness of the quantum mechanical 
path 
of a string. In order to achieve this objective, we need two 
things, a) 
the new form of the uncertainty principle for strings, and b) 
the 
explicit form of the wave--packet for string loops. Then, we 
will be 
able to compute the Hausdorff dimension of a quantum string 
and to 
identify the parameter which controls the transition from the 
smooth 
phase to the fractal phase.\\
There are some finer points of this broadly defined program 
that 
seem worth emphasizing at this introductory stage, before we 
embark on a technical discussion of our results. The main 
point is 
that, unlike superstring theory, our formulation represents 
an 
attempt to construct a {\it quantum mechanical} theory of 
(closed) 
strings in analogy to the familiar case of point--particles. 
The ground 
work of this approach was developed by the authors in two 
previous 
papers, in which we have extended the Hamilton--Jacobi 
formulation 
and Feynman's path integral approach to the case of classical 
and 
quantum closed strings\cite{noi},\cite{prop}. That work was 
largely 
inspired by the line functional approach of Carson and 
Hosotani 
\cite{hoso}, and by the non--canonical quantization method 
proposed 
by Eguchi \cite{egu}, and this is reflected by our 
unconventional 
choice of dynamical variables for the string, namely, the 
spacelike 
area enclosed by the string loop and its 2--form conjugate 
momentum. Furthermore, central to our own quantum mechanical 
approach, is the choice of ~``time variable"~, which we take 
to be the 
timelike, proper area of the string manifold, in analogy to 
the point--particle 
case.\\ 
Presently, we are interested primarily in the analysis of the 
quantum 
fluctuations of a string loop. By quantum fluctuations, we 
mean a 
random transition, or quantum jump, between different string 
configurations. Since in any such process, the {\it shape} of 
the loop 
changes, we refer to it as a ~``shape shifting"~ process. We 
find that 
any such process, random as it is, is subject to an extended 
form of 
the Uncertainty Principle which forbids the exact, 
simultaneous 
knowledge of the string shape and its area conjugate 
momentum. The 
main consequence of the Shape Uncertainty Principle is the 
~``fractalization"~ of the string orbit in spacetime. The 
degree of 
fuzziness of the string world--sheet is measured by its 
Hausdorff 
dimension, whose {\it limiting value} we find to be 
$\displaystyle{D_H=3}$. In order to obtain this result, we 
need the 
gaussian form of a string wave--packet, which we construct as 
an 
explicit solution of the functional Schrodinger equation for 
loops. 
Next, we try to quantify the transition from the classical, 
or smooth 
phase, to the quantum, or fractal phase. Use of the Shape 
Uncertainty 
Principle, and of the explicit form of the loop wave--packet, 
enables 
us to identify the control parameter of the transition with 
the De 
Broglie area characteristic of the loop. Accordingly, the 
paper is 
organized as follows:\\
in Section II, we discuss the basic solutions of the loop 
Schrodinger 
equation. These solutions represent the analogue of the plane 
wave 
and gaussian wave--packet in ordinary quantum mechanics. In 
Section III, we introduce the Shape Uncertainty Principle 
which 
governs the shape shifting processes in loop quantum 
mechanics. 
Section IV is devoted to the fractal properties of the string 
quantum 
path. We compute its Hausdorff dimension and study the 
classical--
to--fractal transition. Section V concludes the paper with a 
summary 
of our results and some final remarks on the formal aspects 
of this 
work.

\section{A functional approach to the Quantum Mechanics of 
closed strings}
	
\subsection{The loop propagator and the loop Schrodinger 
equation}
	A useful  starting point 
        of our discussion is the {\it exact} form of the 
string quantum 
kernel discussed in Ref.[\cite{prop}], namely
        \begin{equation}
        K[C,C_0;A]=\left({m^2\over 2i\pi 
A}\right)^{3/2}\exp\left\{
        {im^2\over 4 A}\left[ \sigma^{\mu\nu}(C)-
\sigma^{\mu\nu}(C_0)
        \right]\left[
        \sigma_{\mu\nu}(C)- 
\sigma_{\mu\nu}(C_0)\right]\right\}\label{k}
        \end{equation}
	where $m^2$ is defined in terms of the string tension, 
$m^2=1/2\pi\alpha'$, $\sigma_{\mu\nu}(C)$
        is the {\it area element} of the loop 
$C:\,x^\mu=x^\mu(s)$ 
\cite{mig},
        i.e.
        \begin{equation}
        \sigma^{\mu\nu}(C)\equiv \oint_C x^\mu dx^\nu
        \end{equation}
	and $A$ is the {\it proper area} of the string manifold, 
invariant under reparametrization of the world--sheet 
coordinates $\{\xi^0,\xi^1\}$
	\begin{equation}
	A={1\over 2}\epsilon_{mn}\int_{\cal S}d\xi^m\wedge 
d\xi^n
	\end{equation}

The geometric doublet ($\sigma_{\mu\nu}, A$) represents the 
set of 
dynamical variables in our formulation of string dynamics. 
This 
choice makes it possible to develop a Hamilton--Jacobi theory 
of 
string loops which represents a natural extension of the 
familiar 
formulation of classical and quantum mechanics of point--
particles 
\cite{noi},\cite{prop}. With hindsight, the analogy becomes 
transparent when one compares Eq.(1) with the
        amplitude for a relativistic particle of mass $m$ to 
propagate from
        $x_0^\mu$ to $x^\mu$ in a proper--time lapse $T$
        \begin{equation}
        K(x ,x_0; T)=\left({m\over 2\pi T}\right)^2
        \exp\left[{i m\over 2 T}\vert x-x_0\vert^2\right] .
        \end{equation}
        This comparison suggests the following correspondence 
between 
dynamical variables for particles and strings:	a particle 
position in  
spacetime is
        labelled by four real numbers $x^\mu$ which represent 
the
	projection of the particle position vector along the
	coordinate axes. In the string case, the conventional 
choice is to 
consider the position vector for each constituent point , and 
then 
follow their individual dynamical evolution in terms of the 
coordinate time $x^0$, or the proper time $\tau$. As a matter 
of fact, 
the canonical string quantization is usually implemented in 
the 
proper time gauge $x^0=\tau$. However, this choice explicitly 
breaks 
the reparametrization invariance of the theory, whereas in 
the 
Hamilton--Jacobi formulation of string dynamics, we have 
insisted 
that reparametrization invariance be manifest at every stage. 
The 
form (1) of the string propagator reflects that requirement. 
We shall 
call $\sigma(C)$ the {\it string configuration tensor} which 
plays the 
same role as the {\it position vector} in the point--particle 
case. 
Indeed, the six components of $\sigma(C)$ represent the 
projection 
of the loop area onto the coordinate planes in spacetime. 
Likewise, 
the reparametrization invariant evolution parameter for the 
string 
turns out to be neither the coordinate time, nor the proper 
time of 
the constituent points, but the proper area $A$ of the {\it 
whole} 
string manifold. As a matter of fact, the string
        world--sheet is the spacetime image, through the 
embedding 
$x^\mu=x^\mu(\sigma^0,
        \sigma^1)$, of a two dimensional manifold of 
coordinates 
$(\sigma^0,
        \sigma^1)$. Thus, just as the proper time $\tau$ is a 
measure of the timelike distance between the final and 
initial position of a point 
particle, the proper area $A$ is a measure of the timelike, 
or 
parametric distance between $C$ and $C_0$, i.e., the final 
and the 
initial configuration of the string. This idea was originally 
proposed 
by Eguchi \cite{egu}.\\	
	It may seem less clear which interior area must be 
assigned to 
any given loop, 
	as one can imagine infinite different surfaces having 
the loop 
as
	a unique boundary. However, once we
	accept the idea to look at the area of a surface $S_C$  
as a sort 
of
	time label for its boundary $\partial S_C$, then the 
	arbitrariness in the assignment of $S_C$ corresponds to 
the 
usual freedom to reset the initial instant of time
	for our clock. The important point is that, once the 
initial area 
has been chosen, the string
	clock measures area lapses.
	In summary, then, $\left[\sigma^{\mu\nu}(C)-
\sigma^{\mu\nu}(C_0)\right]^2$ represents the
         ``~spatial distance squared~'' between $C$ and 
$C_0$, and $A$
	represents the classical time lapse for the string to 
change its 
shape
	from $C_0$ to $C$.\\
		The {\it quantum} formulation of string--dynamics 
based 
on these
non--canonical variables
 was undertaken in \cite{prop} with the evaluation of the 
string 
kernel and the derivation of the Schrodinger loop equation. 
Presently, we are interested in the quantum fluctuations of a 
loop. By 
this, we mean a shape--changing transition, and we would like 
to 
assign a probability amplitude to any such process. In order 
to do 
this, we make use of ~``areal, or loop, derivatives"~, as 
developed, for 
instance, by Migdal \cite{mig}. It may be useful to review 
briefly how 
loop derivatives work, since they
	are often confused with ordinary functional derivatives 
in view 
of the formal relation
	\begin{equation}
	{\delta\over \delta x^\mu(s)}=x^{\prime\nu}(s) 
{\delta\over 
	 \delta \sigma^{\mu\nu}(s)} .
	\end{equation}
	However, there is a basic difference between
	these two types of operation. 
To begin with, an infinitesimal {\it shape variation} 
corresponds to 
``cutting'' the 
	loop $C$ at a particular point, say $y$, and then 
joining the two 
end--points to  an infinitesimal
	loop $\delta C_y$. Accordingly,
	\begin{equation}
        \delta \sigma^{\mu\nu}(C;y)= \oint_{ \delta C_y} 
x^\mu 
dx^\mu\approx
	dy^\mu\wedge dy^\nu \label{darea}
        \end{equation}
	where $dy^\mu\wedge dy^\nu$ is the elementary oriented 
area subtended by $\delta C_y$. A suggestive description of 
this 
procedure, due to Migdal \cite{mig}, 
is that of adding a ~``petal"~ to the original loop. Then, we 
can speak of an 
``~intrinsic distance~''
        between the deformed and initial strings, as the 
infinitesimal,
        oriented area variation,
        $\delta\sigma^{\mu\nu}(C;y)\equiv
        dy^\mu\wedge dy^\nu $. ~``Intrinsic"~, here, means 
that the 
(spacelike) distance is invariant under reparametrization 
and/or
        embedding transformations. Evidently, there is no 
counterpart 
of this operation in the case of point--particles, because of 
the lack of 
spatial extension. Note that, while $\delta x^\mu(s)$ 
represents
	a {\it smooth} deformation of the loop $x^\mu 
=x^\mu(s)$, the 
addition
	of a petal introduces a singular ``{\it cusp}'' at the 
contact
	point.  Moreover, cusps produce 
	infinities in the ordinary variational derivatives but 
not in the
	area derivatives \cite{mig}. \\  
 Non--differentiability is the hallmark of fractal objects. 
Thus, 
anticipating one of our results, quantum loop fluctuations, 
interpreted as singular shape--changing transitions resulting 
from 
~``petal addition"~, are responsible for the fractalization 
of the string. 
Evidently, in order to give substance to this idea, we must 
formulate 
the shape uncertainty principle for loops, and the 
centerpiece of this 
whole discussion becomes the loop  wave functional
$\Psi(C;A)$, whose precise meaning we now wish to discuss.\\
Suppose the shape of the initial string is approximated by 
the loop
configuration $C_0:\,x^\mu=x^\mu_0(s)$. The corresponding 
``wave 
packet''
$\Psi(C_0;0)$ will be concentrated around $C_0$. As the areal 
time 
increases, the initial string evolves, sweeping a world--
sheet of 
parametric proper
area $A$. Once $C_0:\,x^\mu=x^\mu_0(s)$ and $A$ are assigned,
the final string $x^\mu=x^\mu(s)$ can attain any of the 
different
shapes compatible with the given initial condition and with 
the 
extension
of the world--sheet. Each geometric configuration
corresponds to a different ``point'' in loop space 
\cite{prop}. Then, 
$\Psi(C;A)$ {\it
will represent the probability amplitude to find a string of 
shape
$C:\,x^\mu=x^\mu(s)$ as the final boundary of the world--
surface, of
proper area $A$, originating from $C_0$.} From this vantage 
point, 
the
quantum string evolution is a random {\it shape--shifting}
process which corresponds, mathematically, to the spreading 
of the
initial wave packet $\Psi(C_0;0)$ throughout loop space. 
The wavefunctional $\Psi(C;A)$ can be obtained either by 
solving the
loop Schr\"odinger equation

        \begin{equation}
        -{1\over 4m^2}\left(\oint_C dl(s)\right)^{-1}\oint_C 
dl(s)
        {\delta^2\Psi[C;A]\over \delta\sigma^{\mu\nu}(s)
        \delta\sigma_{\mu\nu}(s)}
        =i {\partial \Psi[C;A]\over\partial A}\ .
        \label{nove}
        \end{equation}
	where, $dl(s)\equiv ds\sqrt{x^\prime (s)^2}$ is the 
invariant  
	element of  string length,
or by means of the amplitude (\ref{k}), summing  over all the 
initial 
string
configurations. This amounts to integrate over all the 
allowed loop 
configurations $\sigma(C_0)$:
\begin{eqnarray}
        \Psi[C;A]&=&\sum_{C_0}K[C,C_0;A]\Psi[C_0;0]\nonumber\\
&=&\left({m^2\over 2i\pi A}\right)^{3/2}
\int [{\cal D}\sigma(C_0)]\exp\left[{im^2\over 4 A}\left( 
\sigma(C)-
\sigma(C_0)\right)^2\right]\Psi[C_0;0] .\label{wave}
        \end{eqnarray}\\
	Equation(\ref{nove}), we recall, is the quantum 
transcription, 
through the Correspondence Principle,
	\begin{eqnarray}
	&&H\rightarrow -i {\partial\over\partial A}\label{aham}\\
	&& P_{\mu\nu}(s)\rightarrow i 
	{\delta\over\delta \sigma^{\mu\nu}(s)}\label{def}
	\end{eqnarray}
of the classical relation between the area--hamiltonian $H$ 
and the
	 loop momentum density $P_{\mu\nu}(s)$ \cite{prop} :
	\begin{equation}
	H(C)={1\over 4m^2 }\left(\oint_C dl(s)\right)^{-
1}\oint_C dl(s)
 	P_{\mu\nu}(s)P^{\mu\nu}(s)
	\label{ep}
	\end{equation}
	 
	 Once again, we note the analogy between Eq.(\ref{ep}) 
and the 
familiar energy momentum relation for a point particle, 
$H=p^2/2m$. 
We shall comment on the ``non--relativistic'' form of 
Eq.(\ref{ep}) in 
the concluding section of this paper. Presently, we limit 
ourselves to 
note that the difference between the point--particle case and 
the 
string case, stems from the spatial extension of the loop, 
and is 
reflected in Eq.(\ref{ep}) by the averaging integral of the
	momentum squared along the loop itself. Equation 
(\ref{ep})
	represents the total loop energy instead of the energy 
of a
	single constituent string bit. Just
	as the particle linear momentum gives the direction 
along 
which a 
	particle moves and the rate of position change, so the  
	loop momentum describes the deformation in the loop 
shape 
and
	the rate of shape change. The corresponding hamiltonian 
describes
	the energy variation as the loop area varies, 
irrespective of the
	actual point along the loop where the deformation takes 
place.\\  
		Accordingly, the hamiltonian (\ref{aham}) 
represents the 
{\it generating operator} of the loop area variations, and 
the 
momentum density
	 (\ref{def})  represents the generator of the 
deformations in 
the loop shape at the point $x^\mu(s)$.\\
From the above discussion, we are led to conclude that:\\
	a) deformations may occur randomly at any point on the 
loop;\\
	b) the antisymmetry in the indices $\mu$, $\nu$ 
guarantees 
that
	$P_{\mu\nu}(s)$ generates orthogonal deformations only, 
i.e.
	$P_{\mu\nu}(s)x^{\prime\mu}(s) x^{\prime\nu}(s)\equiv 
0$;\\
 	c) shape changes cost energy because of the string 
tension and 
the fact that, adding a small loop, or ``petal'', increases 
the total
 length of the string;\\
	d) the energy balance condition is provided by equation
	(\ref{ep}) at the classical level and by equation 
(\ref{nove})
	at the quantum level. In both cases the global energy 
variation
	per unit proper area is obtained by a loop average of 
the 
	double deformation at single point.\\
	a), b), c), d) represent the distinctive features of the 
string
	quantum shape shifting phenomenon.

	\subsection{Plane wave solution}
	
	The next step in our program to set up a (spacetime 
covariant) 
functional
	 quantum mechanics of closed strings, is to find the 
basic
	solutions of equation (\ref{nove}). As in the quantum 
mechanics of
	point particles, we seek first plane--wave solutions 
	\begin{eqnarray}
        \Psi(C; 0)=&&{\rm const.}\exp\left({i\over 2}
        \oint_C x^\mu dx^\nu P_{\mu\nu}(x)\right)\nonumber\\
	=&&{\rm const.}\exp\left({i\over 2}\int_0^1 ds\, x^\mu\, 
	{dx^\nu\over ds}P_{\mu\nu}(s)\right) .
	\label{pw}
        \end{eqnarray}
 	The overall constant in the above equation will be fixed 
by a 
suitable normalization
	condition.
	Apart from that, the wave functional (\ref{pw}) 
represents an 
eigenstate
	of the loop total momentum operator
	\begin{equation}
	\left(\oint_C dl(\bar s)\right)^{-1}i\oint_C dl(\bar s)
	{\delta\over\delta \sigma^{\mu\nu}(\bar s)} \Psi(C; 0)=
	P_{\mu\nu}(C)\Psi(C; 0)\label{aread}
	\end{equation}
	with eigenvalues
	\begin{equation}
	 P_{\mu\nu}(C)=\left(\oint_C dl(\bar s)\right)^{-
1}\oint_C 
dl(\bar s)
	P_{\mu\nu}(\bar s)\label{ploop} .
	\end{equation}
	Note that, while $\sigma^{\mu\nu}(C)$ is a {\it 
functional} of 
the loop $C$, i.e., it contains no reference to a special 
point of the 
loop,
	the area derivative $\delta/\delta \sigma^{\mu\nu}(\bar 
s)$ 
operates at the contact point $y^\mu=x^\mu(\bar s)$. Thus, 
the area 
derivative
	of a functional is no longer a functional. Even if the 
functional
	under derivation is reparametrization invariant,  its
	area derivative behaves as a scalar density under 
redefinition 
of the
	loop coordinate. In order to recover reparametrization 
invariance, we have 
	to get rid of the arbitrariness in the choice of the 
attachment 
point.
	 This can be achieved by summing over all its possible
	locations along the loop, and then compensating for the 
overcounting of the
	area variation by averaging the result over the proper 
length 
of the loop. This is what we have done in equations 
(\ref{aread}) and 
(\ref{ploop}), thereby trading
	 the string momentum density $P_{\mu\nu}(\bar s)$, i.e.
	a {\it function} of the loop coordinate $\bar s$, with a 
	{\it functional} loop momentum $P_{\mu\nu}(C)$. The same 
prescription
	enables us to define any other reparametrization 
invariant 
derivative operator.
	Hence, we introduce a simpler but more effective
	 notation, and define the {\it loop derivative} as
	\begin{equation}
	{\delta\over\delta C^{\mu\nu}}\equiv  \left(\oint_C 
dl(s)\right)^{-1}
	\oint_C dl(s){\delta\over\delta 
\sigma^{\mu\nu}(s)}\label{newnotation} .
	\end{equation}
	This represents a genuine loop operation without 
reference to 
the way in which 
	the loop is parametrized.\\

	With the above remarks in mind, the functional Laplacian 
operator appearing in equation (\ref{nove}) is 
reparametrization  
and Lorentz invariant,
and is constructed according to the same prescription: attach 
two 
petals at the loop point $s$, then compute the functional 
variation 
and, finally, 
	average over all possible  locations of the attachment 
point. In
	the notation (\ref{newnotation}), the {\it loop 
laplacian}
	reads
	\begin{equation}
	{1\over 2}{\delta^2\over \delta C^{\mu\nu}\delta 
C_{\mu\nu}}\equiv
	{1\over 2}\left(\oint_C dl(s)\right)^{-1}\oint_C dl(s)
	 {\delta^2\over 
\delta\sigma^{\mu\nu}(s)\delta\sigma_{\mu\nu}(s)} .
	\end{equation}
	Then, if we factorize the explicit $A$-dependence of the 
wave
	functional 	
\begin{equation}
	\Psi(C;A)=\Phi(C)\exp\left(-i{\cal E} A\right)
	\end{equation}
the stationary functional wave equation takes the form 
	\begin{equation}
        -{1\over 4m^2}
        {\delta^2\Psi[C;A]\over \delta C^{\mu\nu}
        \delta C_{\mu\nu}}
        = {\cal E}\, \Psi[C;A]
        \label{stat} .
	\end{equation}
	Thus, we need
	to evaluate the second functional variation of $ \Psi(C; 
0)$
	corresponding to the addition of two petals to $C$, say 
at the 
point 
	$y^\mu=x^\mu(\bar s)$.
 The first variation of $\Psi(C; 0)$
	can be obtained from Eq.(\ref{pw}) 	
\begin{equation}
	\delta \Psi(C; 0)={1\over 2} P_{\mu\nu}(\bar s)
	\delta \sigma^{\mu\nu}(\bar s)\Psi(C; 0)
	\label{vari} .
	\end{equation}
	Reopening the loop $C$ at the same contact point and 
adding a 
second
	infinitesimal loop, we arrive at the second variation of 
	$ \Psi(C; 0)$,	
\begin{equation}
        \delta^2 \Psi(C; 0)={1\over 4} P_{\mu\nu}(\bar 
s)P_{\rho\tau}(\bar s)
        \delta \sigma^{\mu\nu}(\bar s)\delta 
\sigma^{\rho\tau}(\bar s)\Psi(C; 0)
        \label{varii} .
        \end{equation}
	Then, Eq.(\ref{pw}) solves equation (\ref{nove}) if 
	\begin{equation}
	{\cal E}= {1\over 4m^2}\left(\oint_C dl(s)\right)^{-
1}\oint_C dl(s)
	P_{\mu\nu}(s) P^{\mu\nu}(s) ,
	\end{equation}
	which is the classical dispersion relation between 
string energy
	and momentum. Having determined a set of solutions to
	the stationary part of equation (\ref{nove}), the 
complete
	solutions describing the quantum evolution from an 
initial
	state $\Psi(C_0, 0)$ to a final state $\Psi(C, A)$ can 
be obtained
	by means of the  amplitude (\ref{k}), as follows
	\begin{eqnarray}
        \Psi[C;A]&=&\sum_{C_0}K[C,C_0;A]\Psi[C_0;0]\nonumber\\
&=&\left({m^2\over 2i\pi A}\right)^{3/2}
\int [{\cal D}\sigma(C_0)]\exp\left[{im^2\over 4 A}\left( 
\sigma(C)-
\sigma(C_0)\right)^2\right]\Psi[C_0;0] \nonumber\\
&=&{1\over (2\pi)^{3/2}}\exp {i\over 2}\left[\oint_C  
P_{\mu\nu}
        x^\mu dx^\nu-{ A\over 2m^2}\left(\oint_C 
dl(s)\right)^{-1}
        \oint_C dl(s) P_{\mu\nu} P^{\mu\nu}\right]\nonumber\\
        &=&{1\over (2\pi)^{3/2}}\exp {i\over 2}\left[\oint_C  
P_{\mu\nu}
        x^\mu dx^\nu-{\cal E}A\right]
        \label{onda}
	\end{eqnarray}
			
	As one would expect, the solution (\ref{onda}) 
represents a 
``monocromatic
        string wave train'' extending all over loop space.\\

	\subsection{Gaussian Loop--wavepacket}
	
	The quantum state
	represented by Eq.(\ref{pw}) is completely de--localized
        in loop space, which means that all the string shapes 
are
        equally probable, or that the string has no definite 
shape at all.
        Even though the wave functional (\ref{onda}) is a 
solution of
        equation (\ref{nove}), it does not have an immediate 
physical
        interpretation: at most, it can be used to describe a 
flux in
        loop space rather than to describe a single physical 
object.
	Physically acceptable
        one--string states are obtained by a suitable 
superposition of
        ``elementary'' plane wave solutions. The quantum 
state closest 
to a
	classical string will be described by a {\it Gaussian 
wave 
packet},
	\begin{equation}
        \Psi[C_0;0]=\left[{1\over
        2\pi(\Delta\sigma)^2}\right]^{3/4}\exp\left({i\over 
2}
        \oint_{C_0}x^\mu dx^\nu P_{\mu\nu}\right)\exp\left[-
{1\over
        4(\Delta\sigma)^2}\left(\oint_{C_0}x^\mu 
dx^\nu\right)^2\right]\ ,
        \label{gw}
        \end{equation}

	where $\Delta\sigma$ represents the width, or position 
uncertainty in
        loop space, corresponding to an uncertainty in the 
physical 
shape of the loop. By inserting Eq.(\ref{gw}) into 
Eq.(\ref{wave}), and
        integrating out $\sigma^{\mu\nu}(C_0)$, we find

        \begin{eqnarray}
        &&\Psi[C;A]=\left[{1\over 
2\pi(\Delta\sigma)^2}\right]^{3/4}
        {1\over \left(1+i
        A/m^2(\Delta\sigma)^2\right)^{3/2}}\exp\left\{{1\over 
\left(1+i
        A/m^2(\Delta\sigma)^2\right)}\right.\nonumber\\
        &&\left.\left[-{1\over
        4(\Delta\sigma)^2}\sigma^{\mu\nu}(C)\sigma_{\mu\nu}(C)
        +{i\over 2}\oint_C x^\mu dx^\nu P_{\mu\nu}(x)-{i A\over
        4m^2}P_{\mu\nu}(C)
        P^{\mu\nu}(C)\right]\right\}\ .\nonumber\\
        \label{gauss}
        \end{eqnarray}
	
        The wave functional represented by equation 
(\ref{gauss}) spreads
        throughout loop space in conformity with the laws of 
quantum
        mechanics. In particular, the center  of the wave 
packet moves
        according to the stationary phase principle, i.e.

	\begin{equation}
        \sigma^{\mu\nu}(C)-{ A\over m^2}P^{\mu\nu}(C)=0,
        \end{equation}
        and the width broadens as $A$ increases
        \begin{equation}
        \Delta\sigma(A)= \Delta\sigma \left(1+
        A^2/4m^4(\Delta\sigma)^4\right)^{1/2}\ .
        \end{equation}

	Thus, as discussed previously, $A$ represents a measure 
of the 
``time--like''
        distance between the initial and final string loop. 
Then, 
	$m^2(\Delta\sigma)^2$ represent the wavepacket mean 
life.
	As long as $A << m^2(\Delta\sigma)^2$, the wavepacket 
maintains
	its original width $\Delta\sigma$. However, as  $A$ 
increases 
with respect to $m^2(\Delta\sigma)^2$, the
	wavepacket becomes  broader and the initial string
	shape decays in the background space.  
	 Notice that, for sharp  initial wave packets, i.e., for
         $(\Delta\sigma)<< 2\pi\alpha' $,  the shape shifting 
process is
        more ``rapid''
        than for large wave packets. Hence, strings with a 
well defined
        initial shape will sink faster into the sea of 
quantum
        fluctuations than broadly defined string loops.

	\section{The Shape Uncertainty Principle}
	
	In this section we discuss the new form that the 
Uncertainty 
Principle
	takes in the functional theory of string loops.
	Let us consider a Gaussian momentum wave function
	\begin{equation}
	\Phi(P)={1\over \left[\pi(\Delta 
P)^2\right]^{3/4}}\exp\left[
	-{1\over 4(\Delta P)^2}\left(\oint_C dl(s)\right)^{-
1}\oint_C 
dl(s)
	\left(P_{\mu\nu}(s)-K_{\mu\nu}\right)^2\right]\label{gp}
	\end{equation}
	where 
	\begin{equation}
	K_{\mu\nu}(C)={\langle} P_{\mu\nu}(C){\rangle}
	\label{pmedio}
	\end{equation}
	represents the  string momentum mean value around which 
the 
Gaussian wavepacket
	is centered, and
	$\Delta P$ is a measure of the momentum dispersion in 
the 
wavepacket.\\
 	Thus, $K_{\mu\nu}(C)$ is a reparametrization invariant
	{\it functional} of the loop $C$ representing the string 
``drift''
	through loop space.\\
	In order to arrive at the shape uncertainty principle, 
we first 
evaluate the quantum average
	of the loop squared momentum
	\begin{equation}
	{1\over 2}{\langle} 
P_{\mu\nu}(C)P^{\mu\nu}(C){\rangle}= 
	{1\over 2}K_{\mu\nu}(C)K_{\mu\nu}(C)
	+3(\Delta P)^2\label{p2medio} .
	\end{equation}
	Therefore, the string momentum {\it mean square 
deviation}, 
or
	{\it uncertainty} squared, is
	\begin{eqnarray}
	\Delta\Sigma_p^2 &\equiv& {1\over 2}
	{\langle} P_{\mu\nu}(C)P^{\mu\nu}(C){\rangle}-
	{1\over 2}{\langle}P_{\mu\nu}(C){\rangle}
	{\langle}P^{\mu\nu}(C){\rangle}\nonumber\\
	&=& 3(\Delta P)^2\label{deltap} .
	\end{eqnarray}
	The wave functional in configuration space is obtained 
by 
Fourier transforming Eq.(\ref{gp}):
	\begin{eqnarray}
	\Psi(C)&=&{1\over (2\pi)^{3/2}}\int [dP_{\rho\sigma}]
	\Phi(P-K)\exp\left(i\oint_C x^\mu dx^\nu 
P_{\mu\nu}(x)\right)\nonumber\\
	&=&\left[{(\Delta P)^2\over 2\pi}\right]^{3/4}\exp\left[
	-{i\over 2(\Delta P)^2}\oint_C x^\mu dx^\nu K_{\mu\nu}-
	{(\Delta P)^2\over 4}\sigma^{\mu\nu}(C)\sigma_{\mu\nu}
	\right]\label{gs} .
	\end{eqnarray}
	This is again a Gaussian wavepacket, whose ``center of 
mass''
	moves in loop space with a momentum $K_{\mu\nu}$.
	Accordingly, the loop probability density still has a 
Gaussian 
form
	\begin{eqnarray}
	 \vert\Psi(C)\vert^2&=& \left[{(\Delta P)^2\over 
2\pi}\right]^{3/2}
	\exp\left[
	-{(\Delta P)^2\over 
4}\sigma^{\mu\nu}(C)\sigma_{\mu\nu}(C)
	\right]\nonumber\\
	&\equiv& \left[{1\over 
4\pi(\Delta\sigma)^2}\right]^{3/2}\exp\left[
	-{1\over 
4(\Delta\sigma)^2}\sigma^{\mu\nu}(C)\sigma_{\mu\nu}(C)
	\right]\label{densc}
	\end{eqnarray}
	centered around the vanishing loop with a dispersion 
given by
	$\Delta\sigma$. By means of the density (\ref{densc}), 
we 
obtain
	\begin{eqnarray}
	&&{\langle} \sigma^{\mu\nu}(C){\rangle} =0\\
	&&{1\over 2}{\langle} 
\sigma^{\mu\nu}(C)\sigma_{\mu\nu}(C)
	{\rangle}=
	3(\Delta\sigma)^2={3\over 2(\Delta P)^2 }\label{dsigma} 
.
	\end{eqnarray}
	Then, comparing Eq.(\ref{dsigma}) with 
Eq.(\ref{deltap}), we find
	that the uncertainties are related by
	\begin{equation}
	\Delta \Sigma_\sigma \Delta \Sigma_p={3\over \sqrt 2}\ , 
\quad (\hbar=1
	\quad\hbox{units})\label{heis} .
	\end{equation}
	Equation (\ref{heis}) represents the new form that the 
Heisenberg 
	Principle takes when string quantum mechanics is 
formulated in terms
	of diffusion in loop space, or quantum shape shifting. 
Just as a 
	pointlike
	particle cannot have a definite position in space
	and a definite linear momentum at the same time, a 
	physical 
	string cannot have a definite shape and a 
	definite rate of shape changing at a given areal time. 
In other 
words, a string loop cannot be totally at rest neither in 
physical nor
	in loop space: it is subject to a zero--point motion 
characterized by
	\begin{eqnarray}
	&&\langle P_{\mu\nu}(C) \rangle =0\\
	&&\left[{1\over 2}\langle P_{\mu\nu}(C)P^{\mu\nu}(C) 
	\rangle\right]^{1/2}=\sqrt 3(\Delta P)=\Delta\Sigma_p .
	\end{eqnarray}
	In such a state a physical string undergoes a {\it zero-
-point
	shape shifting}, and the loop momentum attains its 
minimum
	value compatible with an area resolution 
$\Delta\sigma$.\\
	To keep ourselves as close as possible to Heisenberg's 
seminal 
idea, we interpret the lack of a definite shape 
as follows:  as we increase the resolution of the 
``microscope''
	used to probe the structure of the string, more and more 
quantum petals
	will appear along the loop. The picture emerging out of 
this is 
	that of a classical line turning into a {\it fractal} 
object as 
	we move from the classical domain of physics to the 
quantum 
	realm of quantum fluctuations. If so,  two questions  
immediately 
	arise:\\
	i) a classical bosonic string is a closed line of
	topological dimension one. Its spacetime image consists 
of a 
	smooth, timelike, two--dimensional world--sheet. Then, 
if a quantum 
	string is a fractal
	object, which Hausdorff dimension should be assigned to 
it?\\
	ii) Is there any {\it critical scale } characterizing 
the
	classical--to--fractal geometrical transition?\\
	These two questions will be addressed in the next 
section.

	\section{ Fractal Strings}

	\subsection{The Hausdorf dimension of a quantum string}

	One of the major achievements of Feynman's formulation 
of quantum
	mechanics was to restore the particle's trajectory
	concept  at the quantum level. 
	However, the dominant contribution to the ``sum over 
histories''
	is provided by trajectories which are nowhere 
differentiable
	\cite{fh}. Non differentiability is the hallmark of 
fractal lines. 
In fact, it seems that Feynman and Hibbs were aware that the 
quantum mechanical path of a particle is inherently fractal. 
This idea was revisited, and further explored by Abbott and 
Wise in the 
case of a free non--relativistic particle \cite{abbw}, 
 	and the extension to relativistic particles was carried 
out by 
several authors, but without a general agreement 
\cite{dhaus}.
	This is one of the reasons for setting up a quantum 
mechanical,
	rather than a field theoretical, framework, even for 
relativistic
	objects. It enables us to adapt the Abbott--Wise 
discussion to 
the string case, with the following basic substitution:
	the point particle, erratically moving through euclidean 
space, 
	is replaced by the string configuration whose 
representative 
	point  randomly 
	drifts  through loop space. We have shown in the 
previous
	sections  that ``flow of time'' for particles is 
replaced by area variations for strings. Hence, the image of 
an abstract linear 
``trajectory'' 
	connecting the two ``points'' $C_0$ and
	$C$ in a lapse of time $A$, corresponds, physically, to 
a family
	of closed lines stacked into a two--dimensional surface 
of
	proper area $A$. Here is where the quantum mechanical 
aspect 
	of our
	approach, and our choice of dynamical variables, seem to 
	have 
	a distinct advantage over the more conventional 
	relativistic description of string dynamics. The 
conventional 
picture of a string world--surface, consisting of a 
collection of 
world--lines associated with each constituent point, is 
replaced by a 
world--sheet ~``foliation"~ consisting of a stack of closed 
lines labelled
	by the internal parameter $A$. In other words, we 
interpret 
the string world--surface as a sequence of ``snapshots''
	of single 
closed lines ordered with respect to the interior area 
bounded by 
them. Then, the randomness of the ``motion'' of a point in 
loop space
	is a reflection of the non--differentiability of the 
string 
world--sheet, which, in turn, is due to zero--point quantum 
fluctuations: the 
random addition of petals to each loop results in a 
fuzziness, or 
graininess of the world surface, by which the string stack 
	 acquires an effective thickness. One can expect that 
this 
graininess becomes apparent only when one can resolve the 
surface 
	small irregularities.\\
The technical discussion on which this picture is based, 
follows 
closely the analysis by Abbott and Wise \cite{abbw}. Thus, 
let us  
divide the string internal coordinates domain into 
	$N$ strips of area $\Delta A$. Accordingly, the string 
stack is
	approximated by the discrete set of the $N+1$ loops, 
$x_n^\mu(s)=x^\mu(s; n\Delta A)\ , n=0,1,\dots N$. 
	Suppose now that we take
	a snapshot of each one of them, and measure their 
internal area.
	If the cross section of the emulsion grains is 
$\Delta\sigma$, 
	then we have  an area indeterminacy $\ge \Delta\sigma$ . 
Then, the 
	total area of the surface subtended by the last and 
	first loop in the stack will be given by
	\begin{equation}
	\langle S \rangle =N \langle \Delta S 
\rangle\label{smedio}
   	\end{equation}
	where $\langle \Delta S \rangle$ is the average area 
variation
	in the interval $\Delta A$,	
\begin{equation}
	\langle \Delta S \rangle\equiv \left[
	\int [d\sigma]\sigma^{\mu\nu}(C)\sigma_{\mu\nu}(C)
	\vert \Psi(C;\Delta A)\vert^2
	\right]^{1/2} .
	\end{equation}
	The finite resolution in $\sigma$ is properly taken into
	account by choosing for $\Psi(C;\Delta A)$ a gaussian 
wave
	functional of the type (\ref{gw}).
	Next, following our correspondence code between 
particles
	and strings, we define the Hausdorff measure ${\cal 
S}_H$ for 
	the stack of fluctuating loops through the relation
	\begin{equation}
	{\cal S}_H =N \langle\Delta S \rangle
	\left(\Delta \sigma\right)^{D_H -2}
	\end{equation}
	where $D_H$ is a number determined by the requirement 
that 
	 $\langle S \rangle $ be independent of 
	the resolution $\Delta \sigma $.\\
	In the above calculation, we may represent the
	quantum state of the string by the loop functional
	
	\begin{equation}
	\Psi(C)=
        \int {[dP_{\mu\nu}(s)]\over
	(2\pi)^{3/2}}\Phi(P)
	\exp {i\over 2}\oint_C P_{\mu\nu} x^\mu dx^\nu
	\end{equation}
 	
	with a gaussian momentum distribution  $\Phi(P)$ 
centered 
around a vanishing string
average momentum $K_{\mu\nu}=0$, i.e., for the moment, we 
consider a free loop
	subject only to zero--point fluctuations. \\
	
	Then,

	\begin{equation}
		\langle \Delta S\rangle\propto {\Delta A\over 4m^2 
\Delta\sigma}
	\sqrt{1+\left({4m^2 (\Delta\sigma)^2\over \Delta 
A}\right)^2},
	\end{equation}

	and, in order to determine the string fractal dimension 
we keep 
$\Delta A$ fixed and take the limit $\Delta\sigma\rightarrow 
0$:

	\begin{equation}
	{\cal S}_H \approx  {N\Delta A\over 4m^2 \Delta\sigma}
	\left(\Delta \sigma\right)^{D_H -2} .
	 \end{equation}
	Hence, in order to eliminate the dependence on 
$\Delta\sigma$, 
	$\displaystyle{D_H=3}$. As in the point particle case, 
quantum 
fluctuations increase by one unit the dimension of the 
	string classical path. As discussed in the following 
subsection, 
the gradual appearance of an extra dimension is perceived as 
fuzziness of the string manifold, and the next question to be 
addressed is which parameter, in our quantum mechanical 
approach, 
controls the transition from classical to fractal geometry.\\

\subsection{Classical--to--fractal geometric transition}

The role of area resolution, as pointed out in the previous 
sections,
leads us to search for a {\it critical area} characterizing 
the transition 
 from classical to fractal geometry of the string stack. 
To this end, we consider tha case in which the string 
possesses a non 
vanishing average momentum, in which case we may use a 
gaussian wave packet of the form (\ref{gp}). The 
corresponding wave functional,  
in the $\sigma$--representation, is

        \begin{equation}
        \Psi_K(C;A)=\
	{[(\Delta\sigma)^2/ 2\pi]^{3/4}\over
   	[(\Delta\sigma)^2+iA/2 m^2]^{3/2}}\exp\left[
	-{\left(\sigma^{\mu\nu}\sigma_{\mu\nu}
	-2i(\Delta\sigma)\sigma^{\mu\nu}K_{\mu\nu}
        +i A(\Delta\sigma)^2 K_{\mu\nu}K^{\mu\nu}/4m^2
	\right)
	\over 2[(\Delta\sigma)^2+iA/2 m^2]}
	\right]   
        \end{equation}
	
	where we have used equation (\ref{dsigma}) to exchange 
	$\Delta P$ with $\Delta\sigma$.
	The corresponding probability density ``evolves'' as 
follows
	\begin{equation}
        \vert\Psi_K(C;A)\vert^2=
	{(2\pi)^{-3/2}\over [(\Delta\sigma)^2+
        A^2/4(\Delta\sigma)^2 m^4]^{3/2}}
	\exp\left[-{\left(\sigma^{\mu\nu}(C)-
	 A  K^{\mu\nu}/ 2(\Delta\sigma) m^2\right)^2
	\over \left[(\Delta\sigma)^2+
        A^2/4(\Delta\sigma)^2m^4\right]}
        \right]
        \label{gaussdens} .
        \end{equation}

	Therefore, the average area variation $\langle \Delta S 
\rangle$, 
	when the loop wave packet drifts with a momentum 
$K_{\mu\nu}(C)$, is
	\begin{equation}
	\langle \Delta S \rangle\equiv \left[
	\int [d\sigma]\sigma^{\mu\nu}(C)\sigma_{\mu\nu}(C)
	\vert \Psi_K(C;\Delta A)\vert^2
	\right]^{1/2}\label{dsk}
	\end{equation}	
	For our purpose, there is no need to compute the exact 
	form of the mean value
	(\ref{dsk}), but only its dependence on $\Delta\sigma$. 
This
	can be done in three steps:\\
	i) introduce the adimensional integration variable
	\begin{equation}
	Y^{\mu\nu}(C)\equiv {\sigma^{\mu\nu}(C)\over 
\Delta\sigma}
 	\end{equation}
	ii) shift the new integration variable as follows
	\begin{equation} 
	Y^{\mu\nu}(C)\rightarrow \bar Y^{\mu\nu}(C)\equiv
	Y^{\mu\nu}(C)-{\Delta A K^{\mu\nu}\over 2m^2 
(\Delta\sigma)^2}
	\end{equation}
	iii) rescale the integration variable as 
	\begin{equation}
	\bar Y^{\mu\nu}(C)\rightarrow
	Z^{\mu\nu}(C)\equiv \bar Y^{\mu\nu}(C)\left[1+
		{(\Delta A )^2\over 4m^4 
(\Delta\sigma)^4}\right]^{1/2} .
	\end{equation}
	Then, we obtain	
\begin{equation}
	\langle \Delta S \rangle=
	{ \Delta A \over \sqrt2\Lambda_{DB}2m^2 (2\pi)^{3/4}}
	\left[\int [dZ]\left(
	{\Lambda_{DB}Z^{\mu\nu}\over 
(\Delta\sigma)}\sqrt{1+\beta^{-2}}+
	{\Lambda_{DB}K^{\mu\nu}}\right)^2e^{-
Z^{\mu\nu}Z_{\mu\nu}/2}
	\right]^{1/2}\label{sclfr}
	\end{equation}
	where,
	\begin{eqnarray}
	&&\Lambda_{DB}^{-1}\equiv 
	\sqrt{{1\over 2}K^{\mu\nu}K_{\mu\nu}}     
\label{dbarea}\\
	&&\beta\equiv {\Delta A\over 2m^2 (\Delta\sigma)^2} .
	\end{eqnarray}
	The parameter $\beta$ measures the ratio of the 
``temporal'' to ``spatial''
	uncertainty, while the area $\Lambda_{DB}$ sets the 
scale of the
	surface variation at which the string momentum is 
$K_{\mu\nu}$. Therefore, always with the particle analogy in 
mind, 
	we shall call $\Lambda_{DB}$ the {\it loop DeBroglie 
area.} 
	Let us assume, for the moment, that $\Delta A$ is 
independent
	of $\Delta\sigma$, so that either quantity can be 
treated as a 
free parameter in the theory. A notable exception to this 
hypothesis
	will be discussed shortly. Presently, we note that 
taking the limit 
	$(\Delta\sigma)\rightarrow 0$, affects only the first 
term of the
	integral (\ref{sclfr}), and that its weight with respect 
to the 
second term is measured by the ratio 
$\Lambda_{DB}/(\Delta\sigma)$. If the area resolution is much 
larger than the loop DeBroglie area, then the first term is 
negligible:  
$\langle \Delta S \rangle$
	is independent of $\Delta\sigma$  
	and $\langle S \rangle$ scales as
	\begin{equation}
	\Lambda_{DB}<< (\Delta\sigma)\ : 
	{\cal S}_H\approx (\Delta\sigma)^{D_H-2}\label{dcl} .
	\end{equation}
	In this case, independence of $(\Delta\sigma)$ is 
achieved by 
assigning $D_H=2$. 
	As one might have anticipated, the detecting apparatus 
is 
unable to resolve the graininess of the string stack, which 
therefore 
appears as a smooth two dimensional surface. \\
	The fractal, or quantum, behavior manifests itself below 
	$\Lambda_{DB}$, 
	when the first term in (\ref{sclfr}) provides the 
leading contribution
	\begin{equation}
        \Lambda_{DB}>> (\Delta\sigma)\ :
        {\cal S}_H\approx {N\Delta A\over \Delta\sigma} 
	(\Delta\sigma)^{D_H-2}
	\sqrt{1+{4m^4 (\Delta\sigma)^4\over (\Delta 
A)^2}}\label{dhq} .
        \end{equation}
	This expression is less transparent than the 
relation(\ref{dcl}), 
as it involves also
	the $\Delta A$ resolution. However, one may now consider 
two 
special subcases in which the Hausdorff dimension can be
	assigned a definite value.\\  
	In the first case, 
	we keep $\Delta A$ fixed, and scale $\Delta\sigma$
	down to zero. Then, each $\langle \Delta S 
\rangle\propto 
	(\Delta\sigma)^{-1}$ diverges, because of larger and 
larger
	shape fluctuations, and 
	\begin{equation}
	{\cal S}_H\approx { A\over \Delta\sigma} 
	(\Delta\sigma)^{D_H-2}
	\end{equation}
	requires $D_H=3$.\\
	The same result can be obtained also in the second 
subcase, in 
which both $\Delta\sigma$ and $\Delta A$ scale down to zero, 
but in 
such a way that their ratio remains constant,	
\begin{equation}
	{2m^2 (\Delta\sigma)^2\over (\Delta 
A)}_{\Delta\sigma\rightarrow 0}
	=const.\equiv {1\over b} .
	\end{equation}
	The total interior area $A=N \Delta A$ is kept fixed. 
Therefore, as
	$\Delta A\sim (\Delta\sigma)^2\rightarrow 0$, 
	then $N\rightarrow \infty$ in order to keep $A$ finite.
	Then,
	\begin{equation}
	\langle \Delta S \rangle\propto
	{\Delta A\over \Delta\sigma}\sqrt{1+{1\over b^2}}
	\propto \Delta\sigma\label{selfsim} ,
	\end{equation}
	and
	\begin{equation}
	{\cal S}_H\propto A (\Delta\sigma)^{D_H-2} 
	{1\over \Delta\sigma }\sqrt{1+{1\over b^2}} ,
	\end{equation}
	which leads to $D_H=3$ again. In the language of fractal 
geometry, this interesting subcase corresponds to 
{\it self--similarity}. Thus, the condition (\ref{selfsim}) 
defines 
a special class of {\it self--similar loops} characterized by 
an average area 
variation which is proportional to $\Delta\sigma$ at any 
scale. \\

\section{Conclusions, some final remarks, and outlook}

Classical string dynamics is based on the simple and 
intuitive notion 
that the world--sheet of a relativistic string consists of a 
{\it smooth, two--dimensional}
manifold embedded in a preexisting spacetime. Switching from 
classical to quantum
dynamics changes this picture in a fundamental way. In the 
path integral approach to particle quantum mechanics, Feynman 
and 
Hibbs were first to point out that the trajectory of a 
particle is 
continuous but nowhere differentiable \cite{fh}. This is 
because, 
according to Heisenberg's principle, when a particle is more 
and more 
precisely located in space, its trajectory becomes more and 
more 
erratic. Abbott and Wise were next to point out that a 
particle 
trajectory, appearing as a smooth line of topological 
dimension one, turns into a {\it fractal} object of 
{\it Hausdorff dimension} two, when the 
resolution of the detecting apparatus is smaller than the 
particle 
De Broglie wavelength\cite{abbw}.  Likewise, extending the 
path 
integral approach to the string case, one must take into 
account  
	the coherent contributions from all 
	the world--sheets satisfying some preassigned boundary 
conditions, and one might expect that a string quantum world-
-sheet 
is a fractal, non--differentiable surface. However, to give a 
{\it 
quantitative} support to this expectation is less immediate 
than it 
might appear at first glance. From an intuitive point of 
view, each
	string bit can be localized only with a finite 
resolution
	in position and momentum. Thus, a string loses its 
classical, 
well defined geometric shape, and its world--sheet appears 
fuzzy as a 
consequence of quantum fluctuations. However, the notion of a  
string 
	world--sheet is fully relativistic, and the 
implementation
	of fractal geometry in high energy physics is an open 
field of 
research with many aspects of it still under discussion 
\cite{dhaus}.
	In this paper, we have argued that it is advantageous to 
work with a first quantized formulation of string dynamics.
	The {\it canonical} quantization approach consists in 
Fourier 
analyzing the string vibrations around an equilibrium 
configuration and  assign
the Fourier coefficients the role of ladder operators 
creating 
and annihilating infinitely many vibration modes. Based on 
our 
previous work \cite{noi},\cite{prop}, we have suggested an 
alternative view of a {\it closed} string quantum vibrations, 
and we readily admit that the exact relationship between the 
two quantization schemes is unclear and requires a more 
exhaustive investigation. However, the conceptual difference 
between the two approaches is sharp: we are quantizing the 
string motion not through the displacement of each point on 
the string, but through the string {\it shape}. The outcome 
of this novel approach is  a {\it quantum
	mechanics of loops,} whose spacetime  interpretation 
involves 
{\it quantum shape shifting transitions}, i.e., 
	quantum mechanical jumps among all possible string 
geometric
	shapes. The emphasis on string shapes, rather than 
points, 
represents  a departure from the canonical formulation and 
requires 
an appropriate choice of dynamical variables, namely, the 
string 
configuration tensor $\sigma(C)$, and the areal time $A$.  
This 
functional approach enables us to extend the {\it quantum 
mechanical} discussion by Abbott and Wise to the case of a 
{\it 
relativistic} (closed) string. It may seem somewhat 
confusing, if not 
contradictory, that we deal with a relativistic system in a 
quantum 
mechanical, i.e., non relativistic framework, as opposed to a 
quantum 
field theoretical framework \cite{mr}. However, this is not 
new in 
theoretical physics \cite{franchi}. In any case, one has to 
realize that 
there are two distinct levels of discussion in our approach. 
At the {\it spacetime level}, where the actual deformations 
in the string 
shape take place, the formulation is fully relativistic, as 
witnessed by the covariant structure of our equations with 
respect to the 
Lorentzian indices. However, at the {\it loop space level}, 
where each 
``point'' is representative of a particular loop 
configuration, our 
formulation is quantum mechanical, in the sense that the 
string coordinates 
$\sigma$ and $A$ are not treated equally, as it is manifest, 
for 
instance, in the loop Schrodinger equation (\ref{nove}). As a 
matter 
of fact, this is the very reason for referring to that 
equation as the 
~``Schrodinger equation"~ of string dynamics: the timelike 
variable 
$A$ enters the equation through a first order partial 
derivative, as 
opposed to the functional ~``laplacian"~, which is of second 
order with 
respect to the spacelike variables $\sigma$. Far from being 
an 
artifact of our formulation, we emphasize that this spacetime 
covariant, quantum mechanics of loops, is a direct 
consequence of the 
Hamilton--Jacobi formulation of classical string dynamics. 
This is 
especially evident in the form of the classical area--
Hamiltonian, 
equation (\ref{ep}), from which, via the correspondence 
principle, we 
have derived the loop Schrodinger equation.\\
Incidentally, our formulation raises the interesting question  
as to 
whether it is possible to ~``covariantize"~ the Schrodinger 
equation in 
{\it loop space}, i.e., to treat the spacelike and timelike 
generalized 
coordinates on the same footing. At the moment this is an 
open 
problem. However, if history is any guide, one might think of 
generalizing the loop Shrodinger equation into a functional 
Klein--Gordon equation, or, to follow Dirac's pioneering work 
and take the 
~``square root"~ of the functional laplacian in order to 
arrive at a first 
order functional equation for string loops. This second route 
would 
involve an extension of Clifford's algebra along the lines 
suggested, for instance, by Hosotani in the case of a 
membrane 
\cite{hoso}.
Be that as it may, with our present quantum mechanical 
formulation, 
we have given a concrete meaning to the fractalization of a 
string 
orbit's in spacetime in terms of the shape uncertainty 
principle. We have concluded that the Hausdorff dimension of 
a quantum 
string's world--surface is three, and that two distinct 
geometric 
phases exist above and below the loop De Broglie area. We 
must emphasize 
that $D_H=3$ represents a limiting value of the Hausdorff 
dimension, in the sense specified in Section IV. In actual 
fact, the 
world--surface of 
a quantum string is literally ~``fuzzy"~ to a degree which 
depends 
critically on the parameter $\beta$, the ratio between 
temporal and 
spatial resolution. Self similarity, we have shown, 
corresponds to a 
constant value of $\beta$, with $D_H=3$. In such a case, the 
shape 
shifting fluctuations generated by petal addition, 
effectively give rise 
to a full transverse dimension in the string stack.\\
As a final remark, we note that the quantum mechanical 
approach 
discussed in this paper is in no way restricted to string--
like objects. 
In principle, it can be extended to any quantum p--brane, and 
we 
anticipate that the limiting value of the corresponding 
fractal 
dimension would be $D_H=p+2$. Then, if the above over all 
picture is 
correct, p--brane fuzziness not only acquires a well defined 
meaning, 
but points to a fundamental change in our perception of 
physical 
spacetime. Far from being a smooth, four--dimensional 
manifold 
assigned ~``ab initio"~, spacetime is, rather, a ~``process 
in the 
making"~, showing an ever changing fractal structure which 
responds 
dynamically to the resolving power of the detecting 
apparatus.


\begin{thebibliography}{99}
\bibitem{witten}Edward Witten,
        Physics Today,24, April 1996	
\bibitem{noi}  A. Aurilia, E.Spallucci, I.Vanzetta, 
        Phys.\ Rev.\ {\bf D50}, 6490, (1994)
 \bibitem{prop} S.Ansoldi, A. Aurilia, E.Spallucci, 
        Phys.\ Rev.\ {\bf D53}, 870, (1996)
\bibitem{hoso} L. Carson, Y. Hosotani, 
	 Phys.\ Rev.\ {\bf D37}, 1492, (1988)
\bibitem{egu} T.Eguchi, 
	 Phys.\ Rev.\ Lett.\ {\bf 44}, 126, (1980)
 \bibitem{mig} A.M.Polyakov, 
	Nucl.\ Phys.\ {\bf B164}, 171, (1980)\\
	A.A. Migdal, 
         Phys.\ Rep.\ {\bf C 102}, 199, (1983)
\bibitem{fh} R.P.Feynman, A.R.Hibbs, 
        {\it Quantum Mechanics and Path Integrals}, Mc Graw--
Hill, N.Y., 
(1965)
\bibitem{abbw} L.F.Abbott, M.B.Wise, 
        Am.\ J.\ Phys.\ {\bf 49}, 37, (1981)
\bibitem{dhaus} G.N.Ord, 
        J.\ Phys.\ A\ {\bf 16}, 1869 (1983)\\
        F.Cannata, L.Ferrari, 
        Am.\ J.\ Phys. {\bf 56}, 721 (1988)\\
        L. Nottale, 
        Int.\ J.\ Mod.\ Phys.\ A\ {\bf 4}, 5047 (1989)\\
        L. Nottale, 
	{\it Fractal Spacetime and Microphysics,} World 
Scientific, 
(1992) 			       	        	
\bibitem{mr} C. Marshall, P. Ramond, 
	Nucl.\ Phys.\ {\bf B85}, 375, (1975)
\bibitem{franchi} A. Kyprianidis, 
	Phys.\ Rep.\ {\bf 155}, 1, (1987)\\
	J.R.Franchi, 
	Found.\ of\ Phys.\ {\bf 23}, 487, (1993)
	\bibitem{hoso} L.Carson, Y.Hosotani
	 Phys.\ Rev.\ {\bf D37}, 1492, (1988)\\
	C-Lin Ho, L.Carson, Y.Hosotani
         Phys.\ Rev.\ {\bf D37}, 1519, (1988)\\
        	\end{thebibliography}
	\end{document}